\documentclass[onecolumn, a4size, 11pt,final]{IEEEtran}
\usepackage{amsmath}
\usepackage{amssymb}
\usepackage{amsfonts}
\usepackage[final]{graphicx}
\usepackage{ctable}
\usepackage{algorithm}
\usepackage{algorithmic}

 \title{Multi-antenna Wireless Powered Communication with Co-channel Energy and Information Transfer}
 \author{Yueling Che,~\IEEEmembership{Member,~IEEE,}  Jie Xu,~{\it Member,~IEEE,} 
 Lingjie Duan,~\IEEEmembership{Member,~IEEE,} \\and Rui Zhang,~\IEEEmembership{Senior Member,~IEEE}  
 \thanks{Y. L. Che, J. Xu, and L. Duan are with the Engineering Systems and Design Pillar,
Singapore University of Technology and Design (e-mail:
\{yueling\_che, jie\_xu, lingjie\_duan\}@sutd.edu.sg). J. Xu is the corresponding author.}
 \thanks{R.~Zhang is with the Department of Electrical and Computer Engineering,
 National University of Singapore (e-mail: elezhang@nus.edu.sg). He is also with
 the Institute for Infocomm Research, A*STAR, Singapore.}
 }

\begin{document}

\maketitle
\thispagestyle{empty}

\begin{abstract}
This letter studies a multi-antenna wireless powered communication (WPC) system with co-channel energy and information transfer, where a wireless device (WD), powered up by wireless energy transfer (WET) from an energy transmitter (ET), communicates to an information receiver (IR) over the same frequency band. We  maximize the achievable data rate from the WD to the IR by jointly optimizing the energy beamforming at the ET and the information beamforming at the WD, subject to their individual transmit power constraints. We obtain the optimal solution to this problem in closed-form, where   the optimal energy beamforming  at the ET achieves a best energy/interference tradeoff between maximizing the energy transfer efficiency to the WD and minimizing the  co-channel interference to the IR. Numerical results show that our proposed  optimal co-channel design is superior to other reference schemes.

\end{abstract}


\newtheorem{definition}{\underline{Definition}}[section]
\newtheorem{fact}{Fact}
\newtheorem{assumption}{Assumption}
\newtheorem{theorem}{\underline{Theorem}}[section]
\newtheorem{lemma}{\underline{Lemma}}[section]
\newtheorem{corollary}{\underline{Corollary}}[section]
\newtheorem{proposition}{\underline{Proposition}}[section]
\newtheorem{example}{\underline{Example}}[section]
\newtheorem{remark}{\underline{Remark}}[section]
\newcommand{\mv}[1]{\mbox{\boldmath{$ #1 $}}}
\newtheorem{property}{\underline{Property}}[section]

\section{Introduction}

Radio-frequency (RF) signals enabled wireless energy transfer (WET) has been recognized as a promising technology to provide  perpetual and convenient power supply to energy-constrained wireless networks.  This motivates an appealing wireless powered communication (WPC) system, in which   wireless information transfer (WIT) from  wireless devices (WDs)  to    information receivers  (IRs), e.g.,   sensor nodes delivering information to fusion centers,   is powered up  by the means of WET from  dedicatedly deployed energy transmitters (ETs). In practice, the IR  and the ET  can be either  separately located as two nodes (e.g., an information access point and a  power beacon) \cite{Huang.TWC.14}-\cite{Liu.Secure}, or co-located as a single node (e.g., a hybrid access point)   \cite{Zhou.TWC.sub}-\cite{Liu.TWC.14}. In both    cases,   WET and WIT are usually assumed to be implemented over orthogonal time/frequency resources \cite{Huang.TWC.14}-\cite{Liu.TWC.14}, so as to avoid  co-channel interference from WET 
to WIT links at a cost of reduced spectrum utilization efficiency.

\begin{figure}
\centering
\DeclareGraphicsExtensions{.eps,.mps,.pdf,.jpg,.png}
\DeclareGraphicsRule{*}{eps}{*}{}
\includegraphics[angle=0,width=0.5\textwidth]{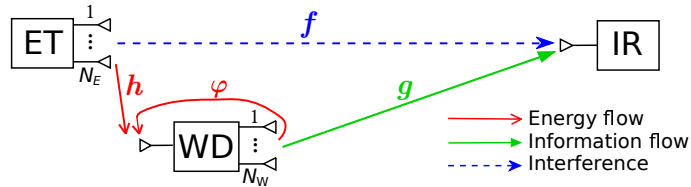}
\caption{ WPC system model with co-channel energy and information transfer.}
\label{fig: system_model}
\end{figure}

To improve the spectrum utilization efficiency, in this letter, we propose a new co-channel energy and information transfer scheme in a WPC system with separately located IR and ET, as  shown in Fig.~\ref{fig: system_model}, where  the WET and WIT links  concurrently utilize the same frequency band, and the WD operates in a full-duplex mode with concurrent energy harvesting and information transmission.
In this case, the co-channel interference from WET  to  WIT links is a key issue that limits the system performance. Therefore, it is essential for the ET to design its transmitted energy signal to balance the tradeoff between maximizing the energy transfer efficiency to the WD and minimizing the co-channel interference to the IR. On  the other hand, the WD also needs to efficiently utilize its harvested energy from the ET  and the recycled self-energy due to the full-duplex operation,  so as to assure its information transmission quality to the IR. To this end, we apply  multi-antenna techniques by assuming the ET and WD are both deployed with multiple transmit antennas.

Under this setup, in this letter, we  maximize the achievable data rate from the WD to the IR, by jointly optimizing the transmit energy signal at the ET and the transmit information signal at the WD  subject to their individual transmit power constraints. We obtain the optimal solution to this problem in closed-form, where   the obtained energy beamforming  at the ET optimally balances the aforementioned energy/interference tradeoff. It is also revealed that the optimal energy and information beamforming can be implemented at the ET and the WD in a distributed manner by only using their respective local channel state information (CSI).
Numerical  results  show that the proposed optimal co-channel design achieves large throughput gains over  three reference schemes.
%including two heuristic co-channel schemes each of which   maximizes  the harvested energy at the WD and cancels the resultant interference at the IR from the ET, respectively,  as well as   a time-division based orthogonal energy and information transfer scheme.

In the literature, there are only  limited studies considering multi-antenna co-channel  WET and WIT \cite{Xu.letter.15} and \cite{Zeng.letter.15}. However,  \cite{Xu.letter.15} studied the coexistence  of two separate WET and WIT systems, instead of a single WPC system as in this paper, while   \cite{Zeng.letter.15}   ignored the co-channel interference from  WET to  WIT links. To our best knowledge, our results on the multi-antenna WPC system with   co-channel energy and information transfer  are new and  have not been reported in the literature.

 \begin{figure}
\centering
\DeclareGraphicsExtensions{.eps,.mps,.pdf,.jpg,.png}
\DeclareGraphicsRule{*}{eps}{*}{}
\includegraphics[angle=0,width=0.7\textwidth]{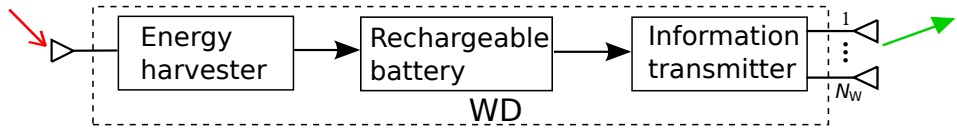}
\caption{A full-duplex wireless powered WD with concurrent energy harvesting and information transmission.}
\label{fig: WD}
\end{figure}

\section{System Model}

We consider a WPC system consisting of one ET, one WD, and one IR, as shown in Fig.~\ref{fig: system_model}, where  the ET delivers wireless   energy  to the WD,  and at the same time  the WD  communicates to the IR over the same channel, by using the harvested energy from the ET and the recycled self-energy from its own transmission.
We assume that the ET is equipped with  $N_E\!>\!1$ antennas for energy transmission,  and  the IR is deployed with a single antenna for information reception. Moreover, the  WD is  deployed with   $1+N_W$ antennas, $N_W\!>\!1$, {with one  for energy harvesting and the other $N_W$ for   information  transmission.\footnote{Note that using multiple receive antennas  for improving the energy harvesting efficiency is equivalent to using a single receive antenna with a larger aperture area. In practice, it is cost-effective for the WD to install a single receive antenna or rectenna for harvesting the RF energy.} In practice, to simultaneously harvest energy and transmit information, the WD can use a full-duplex rechargeable battery that can be charged and discharged at the same time \cite{Huang.TWC.14}, \cite{Zeng.letter.15}, an example of which is shown in Fig.~\ref{fig: WD}.  
Denote the baseband equivalent channels from the ET to the WD and the IR  by vectors  $\mv h \!\in \! \mathbb{C}^{N_E \times 1}$ and $\mv f \!\in \! \mathbb{C}^{N_E \times 1}$, respectively, and that from  the WD to the IR  by $\mv g \!\in \! \mathbb{C}^{N_W \times 1}$.   Also denote the loop channel from the $N_W$ transmit antennas of  the WD to its receive antenna    by  $\mv \varphi \!\in \! \mathbb{C}^{N_W \times 1}$.   It is assumed that all channels are quasi-static flat-fading, where
$\mv h$, $\mv  f$, $\mv g$, and $\mv \varphi$ remain constant during each transmission block with a  length of $T\!>\!0$, but can vary from one block to another.
It is further assumed that the ET perfectly knows the CSI of $\mv h$ and $\mv f$,  while the WD  accurately knows  $\mv g$ and $\mv \varphi$.\footnote{CSI acquisition at the ET and the WD may practically require the corresponding (energy/information) receivers to feed back certain information (see, e.g., \cite{Xu.TSP.14}), and there may exist   CSI imperfections due to, e.g., feedback errors. However, the detailed CSI acquisition methods and the effects of CSI imperfections are beyond the scope of this letter.}
We denote  the energy signal transmitted by  the ET as $\mv x_E \!\in \!\mathbb{C}^{N_E \times 1}$ and the information signal  transmitted by  the WD as   $\mv x_I\!\in\! \mathbb{C}^{N_W \times 1}$.

Consider first the WET from the ET to the WD. We consider that the  transmitted energy signal $\mv x_E$  by the ET is a circularly symmetric complex Gaussian (CSCG) random sequence  to meet certain power spectral density (PSD) requirements set by spectrum regulators.  Accordingly, we denote the energy covariance matrix at the ET as  $\mv Q \!=\! \mathbb{E}\left(\mv x_E \mv x_E^H \right) \!\succeq\! \mv 0$, where $\mathbb{E}(\cdot)$ denotes the statistical expectation, the superscript $H$ denotes the conjugate transpose, and ${\mv A} \succeq {\mv 0}$ means that $\mv A $ is positive semi-definite.     Here, we consider that $\mv Q$ is of a general rank without loss of generality \cite{Xu.TSP.14}.
Denote the maximum transmit power at the ET  by $P$. The transmit power constraint at the ET is given by
\begin{equation}
\mathbb{E}\left(\|\mv x_E\|^2\right) = \mathrm{tr}(\mv Q) \le P, \label{eq: ET_constraint}
\end{equation}
where $\mathrm{tr}(\mv A)$ denotes the trace of a square matrix $\mv A$.

Next, consider   the co-channel WIT from the WD to the IR. We consider transmit information beamforming at the WD. Thus, the  information signal $\mv x_I$ transmitted by the WD can be expressed as  $\mv x_I \!=\! \mv w s$, where $\mv w\! \in \!\mathbb{C}^{N_W \times 1}$ denotes the information beamforming vector  at the WD and $s$ denotes the desired signal for the IR.
We assume that $s$ is a   CSCG random variable with zero mean and unit variance,  and is independent of the energy signal $\mv x_E$ sent by the ET.
Due to the co-channel energy and information transfer,  the received signal at the IR is $y  =  \mv g^H \mv w s  +  \mv f^H \mv x_E  +  n$,
where $\mv g^H \mv w s$ is the  desirable information signal,    $\mv f^H \mv x_E$ is the  interference due to the energy signal sent from the ET,  and $n$ is the additive white Gaussian noise (AWGN)  with zero mean and variance $\sigma^2$.
Hence, the received signal-to-interference-plus-noise ratio (SINR)   at the IR is given by $\gamma \!= \!\left|\mv g^H \mv w\right|^2\!\!\! \Big/\!\left(\mv f^H\mv Q \mv f \!+\! \sigma^2\right)$, where   $\mv f^H\mv Q \mv f$ is the received  interference power at the IR due to the ET's energy transfer. Therefore,  the achievable data rate from the WD to the IR (in bps/Hz) is expressed as 
\begin{equation}
R=\log_2(1+\gamma)=\log_2\left(1+\frac{\left|\mv g^H \mv w\right|^2}{\mv f^H\mv Q \mv f + \sigma^2}\right). \label{eq: SINR}
\end{equation}

Finally, consider the energy harvesting at the WD. The WD can harvest the wireless energy from the ET and also recycle its self-energy via the loop channel.
The totally harvested energy at the WD is thus expressed as  
\begin{equation}
E=\eta\mathbb{E}\left(|\mv h^H \mv x_E+ \mv \varphi^H \mv w s|^2\right)=\eta \mv h^H \mv Q \mv h+\eta \left|\mv \varphi^H \mv w \right|^2, \label{eq: available_energy}
\end{equation}
where $\eta\in(0,1]$ is a constant denoting the energy harvesting efficiency at the WD.
Since the transmit energy at the WD cannot exceed its totally harvested energy  $E$ in (\ref{eq: available_energy}), we obtain the energy harvesting constraint at the WD as  
\begin{equation}
\mathbb{E}(\|\mv x_I\|^2) = \left\|\mv w\right\|^2 \le \eta \mv h^H \mv Q \mv h+\eta \left|\mv \varphi^H \mv w \right|^2,
\end{equation}
or equivalently, 
\begin{equation}
 \mv w^H (\mv I - \eta\mv \varphi \mv\varphi^H) \mv w \le \eta\mv h^H \mv Q\mv h. \label{eq: WD_constraint}
\end{equation}

\section{Optimal  Co-channel Design}
This section studies the optimal co-channel transmission design.   In particular,   we aim to maximize the achievable data rate $R $ from the WD to the IR in (\ref{eq: SINR}), by jointly optimizing   the  ET's   energy  covariance matrix $\mv Q$ and the WD's   information beamforming vector $\mv w$, subject to the  transmit power constraint at the ET in (\ref{eq: ET_constraint}) and the energy harvesting constraint at the WD in (\ref{eq: WD_constraint}). Therefore, the optimization problem is formulated as  
\begin{align}
{\textrm{(P1)}}:~\max_{\mv Q\succeq \mv 0,\mv w} ~& \log_2\left(1+\frac{\left|\mv g^H \mv w\right|^2}{\mv f^H\mv Q \mv f + \sigma^2}\right)\nonumber \\
\mathrm{s.t.}~& \textrm{(1)} \textrm{~and~} \textrm{(5)}. \nonumber
\end{align}

Note that since the objective of (P1) is non-convex, problem (P1) is non-convex in general. Despite this fact, we can still obtain the optimal solution to (P1) by first   deriving the optimal $\mv w$ under any given $\mv Q$, and then obtaining  the optimal $\mv Q$.

First, we optimize $\mv w$ under any given $\mv Q$. It is observed that the objective function of (P1) is monotonically increasing over $\left|\mv g^H \mv w\right|^2$. Optimizing $\mv w$ under a given $\mv Q$ is thus equivalent to solving the following problem. 
\begin{align}
{\textrm{(P2)}}:~\max_{\mv w} ~& \left|\mv g^H \mv w\right|^2,  \nonumber \\
\mathrm{s.t.}~& \textrm{(5)}. \nonumber
\end{align}
To solve (P2), we define  $\tilde{\mv w}\!\triangleq\! \left(\mv I-\eta \mv \varphi \mv \varphi^H \right)^{\!1/2}\!\mv w$, and thus $\mv w\!=\!\! \left(\mv I\!-\!\eta \mv \varphi \mv \varphi^H \right)^{\!-1/2}\!\tilde{\mv w}$, where  $\mv I-\eta \mv \varphi \mv \varphi^H \!\succeq \!\mv 0$ holds practically since $\mv \varphi$ is the loop channel vector with high attenuation.  By replacing $\mv w$ as $\left(\mv I\!-\!\eta \mv \varphi \mv \varphi^H \right)^{\!-1/2}\!\!\tilde{\mv w}$,  (P2) can be rewritten as 
\begin{align}
{\textrm{(P2-E)}}:~\max_{\tilde{\mv w}}  ~&  \left|\mv g^H \! \!\left(\mv I\!-\!\eta \mv \varphi \mv \varphi^H \right)^{\!-1/2\!}\tilde{\mv w}\right|^2\!, \nonumber \\
\mathrm{s.t.}~& \left\|\tilde{\mv w}\right\|^2 \!\le \eta \mv h^H \mv Q \mv h. \nonumber
\end{align}
It is easy to verify that for any arbitrary $\mv Q$, the optimal $\tilde{\mv w}$ for problem (P2-E), denoted by $\tilde{\mv w}(\mv Q)$,    is obtained as
$\tilde{\mv w}(\mv Q)=\sqrt{\eta \mv h^H \mv Q \mv h}\frac{\tilde{\mv g}}{\|\tilde{\mv g}\|}$, where $\tilde{\mv g}=\left(\mv I\!-\!\eta \mv \varphi \mv \varphi^H \right)^{-1/2} \mv g$. As a result, for any arbitrary $\mv Q$,  the optimal $\mv w$ for problem (P2), denoted by $\mv w(\mv Q) $, is obtained as 
\begin{align}
 \mv w(\mv Q) &=   \left(\mv I-\eta \mv \varphi \mv \varphi^H \right)^{-1/2}\tilde{\mv w}(\mv Q) \nonumber \\
 &=\sqrt{\eta \mv h^H \mv Q \mv h}\frac{\left(\mv I\!-\!\eta \mv \varphi \mv \varphi^H \right)^{-1}  \mv g}{\left\|\left(\mv I\!-\!\eta \mv \varphi \mv \varphi^H \right)^{-1/2} \mv g\right\|}. \label{eq: optimal_w_Q}
\end{align}

Next, with the optimal $\mv w(\mv Q)$ for any given $\mv Q$ at hand, we derive the optimal $\mv Q$ for problem (P1). By substituting $\mv w(\mv Q)$ in  (\ref{eq: optimal_w_Q}) into (P1), the optimization of $\mv Q$ becomes  solving the following problem. 
\begin{align}
{\textrm{(P3)}}:~\max_{\mv Q\succeq \mv 0}~& \log_2\!\left(\!1+\eta \|\tilde{\mv g}\|^2\frac{\mv h^H \mv Q \mv h}{\mv f^H\mv Q \mv f + \sigma^2}\right),\nonumber \\
\mathrm{s.t.}~& \textrm{(1)}. \nonumber
\end{align}
The optimal solution to problem (P3), denoted by $\mv Q^{*}$, is given in the following proposition.
\begin{proposition} \label{proposition: optimal_Q}
Let $\mv v=\left( \mv f \mv f^H+\frac{\sigma^2}{P}\mv I\right)^{-1} \mv h$. The optimal $\mv Q^{*}$ for problem (P3) and thus (P1) is
\begin{equation}
 {\mv Q}^{*} = \frac{P \mv v \mv v^H}{\|\mv v \|^2} \label{eq: optimal_Q}
\end{equation}
\end{proposition}
\begin{IEEEproof}
 Please refer to Appendix A.
\end{IEEEproof}
It follows from Proposition \ref{proposition: optimal_Q} that  $\mathrm{rank}(\mv Q^*) = 1$, which indicates that a single energy beam is sufficient for the ET to achieve the optimality.
It is also observed from (\ref{eq: optimal_Q}) that   the optimal energy beamforming at the ET   balances the energy/interference tradeoff, between  maximizing the energy transfer efficiency over the WET link $\mv h$ and minimizing the   resulted co-channel interference over the interference link $\mv f$.

Finally, by substituting  $\mv Q^*$ into $\mv w(\mv Q)$ given in (\ref{eq: optimal_w_Q}), we obtain the optimal $\mv w^*$ for problem (P1) as  
\begin{equation}
 \mv w^*=\mv w(\mv Q^*)=\sqrt{\eta \mv h^H \mv Q^* \mv h}\frac{\left(\mv I\!-\!\eta \mv \varphi \mv \varphi^H \right)^{-1}  \mv g}{\left\|\left(\mv I\!-\!\eta \mv \varphi \mv \varphi^H \right)^{-1/2} \mv g\right\|}. \label{eq: optimal_w}
\end{equation}
By combining $\mv Q^*$ in (\ref{eq: optimal_Q}) and $ \mv w^*$ in (\ref{eq: optimal_w}), the optimal solution to problem (P1) is obtained.
\begin{remark}[distributed implementation]
From  (\ref{eq: optimal_Q}), the optimal $\mv Q^*$  only depends on the ET's local CSI of $\mv h$ and $\mv f$, and as a result,  the optimal energy beamforming can be implemented at the ET locally. Nevertheless, it is observed from (\ref{eq: optimal_w})  that the optimal information beamforming vector $\mv w^*$ generally relies on $\mv h$, $\mv f$,  $\mv g$, and $\mv \varphi$. Despite this fact, $\mv w^*$ can still be obtained by the WD with its local CSI of $\mv g$ and $\mv \varphi$, by noticing that    $\eta \mv h^H \mv Q^* \mv h$ is   the WD's harvested energy from the ET and thus can   be measured by the WD locally.  Therefore, the optimal joint design of $\mv Q^*$ and $\mv w^*$ can be implemented at the ET and the WD in a distributed manner.
\end{remark}

\section{Reference Schemes}
This section considers three reference schemes for performance comparison. First, under the co-channel energy and information transfer, we develop two heuristic schemes for the ET to maximize its transferred energy to the WD or cancel its resultant interference to the IR, respectively. Next, we consider a  time-division based orthogonal  transfer scheme.

\subsubsection{Co-channel Scheme with Harvested Energy Maximization}
In this scheme  with co-channel energy and information transfer, we first optimize the ET's  energy covariance matrix $\mv Q$ to maximize its transferred energy $E$  to the WD in (\ref{eq: available_energy}) subject to the transmit power constraint in (\ref{eq: ET_constraint}), for which we have the  optimal energy covariance matrix at the ET as $\mv Q_E^{*}\!=\!\frac{P \mv h \mv h^H}{\|\mv h\|^2}$.
Next,  under the optimal $\mv Q_E^{*}$, we  optimize the WD's information beamforming vector $\mv w$ to maximize its  data rate $R$ in (\ref{eq: SINR}) subject to the energy harvesting constraint in (\ref{eq: WD_constraint}), for which we obtain the  optimal information beamforming vector at the WD as $\mv w_E^*\!=\!\mv w(\mv Q_E^*)$ with $\mv w(\mv Q)$ given in (\ref{eq: optimal_w_Q}).
\begin{remark}
It can be shown that the optimal   $\mv Q_E^*$ and $\mv w_E^*$ in this reference scheme is   the optimal solution   $\mv Q^*$ and $\mv w^*$ to problem (P1) in the special case when $P\rightarrow 0$.
\end{remark}

\subsubsection{Co-channel Scheme with Interference Nulling}
We also consider  co-channel energy and information transfer in this scheme. First, we optimize the ET's  energy covariance matrix $\mv Q$ to  maximize its transferred energy $E$ to the WD  in (\ref{eq: available_energy}), under the ET's transmit power constraint in (\ref{eq: ET_constraint}) and the zero-interference constraint $\mv f^H \! \mv Q \mv f=0$,  i.e.,  
\begin{align}
{\textrm{(P4)}}:~\max_{\mv Q \succeq \mv 0}~& \mv h ^H \mv Q \mv h,\nonumber \\
\mathrm{s.t.}~&  \textrm{(1)} \textrm{~and~}\mv f^H \! \mv Q \mv f=0.  \nonumber
\end{align}

Let $\mv u=\left(\!\mv I\!-\! \mv f \mv f^H\!\big/ \!\left\|\mv f\right\|^2 \right)\left(\!\mv I\!-\! \mv f \mv f^H\!\big/ \!\left\|\mv f\right\|^2 \right)^H\mv h$.
It can be shown that the optimal $\mv Q$ to problem (P4)  is given by $\mv Q_{IC}^{*}=\frac{P \mv u\mv u^H}{\|\mv u \|^2} $.
Next,  under the optimal  $\mv Q_{IC}^{*}$,  we derive the  WD's optimal information beamforming vector of this scheme, denoted by $\mv w_{IC}^*$, to  maximize the WD's   data rate $R$ in (\ref{eq: SINR})  subject to the energy harvesting constraint in (\ref{eq: WD_constraint}), for which we obtain $\mv w_{IC}^*=\mv w(\mv Q_{IC}^*)$ with $\mv w(\mv Q)$ given in (\ref{eq: optimal_w_Q}).
\begin{remark}
The optimal $\mv Q_{IC}^*$ and $\mv w_{IC}^*$ to problem (P4) can be shown to be the optimal solution   $\mv Q^*$ and $\mv w^{*}$ to problem (P1) in the special case when $P\rightarrow \infty$.
\end{remark}

\subsubsection{Time-Division based Orthogonal Scheme}
 \begin{figure}
\centering
\DeclareGraphicsExtensions{.eps,.mps,.pdf,.jpg,.png}
\DeclareGraphicsRule{*}{eps}{*}{}
\includegraphics[angle=0,width=0.7\textwidth]{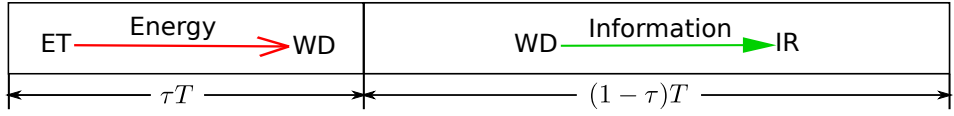}
\caption{Time-division based orthogonal energy and information transfer.}
\label{fig: orthogonal}
\end{figure}
For the time-division based orthogonal transfer, 
  as shown in Fig.~\ref{fig: orthogonal},  each block  of duration $T$ is divided  into two phases:   the first phase of duration $\tau T$, $\tau\in(0,1)$, is used for energy transfer from the ET to the WD, and the second phase of duration $(1-\tau)T$ is for information transfer from the WD to the IR.    For any information beamforming vector $\mv w$, the WD's consumed energy during the information transfer phase is given by $(1-\tau)T\left\|\mv w\right\|^2$, which cannot exceed its harvested energy during the energy transfer phase, given by  $\tau T \eta \mv h ^H \mv Q \mv h$. Hence, we have $\left\|\mv w\right\|^2 \!\!\le\! \eta \frac{\tau}{1\!-\!\tau} \mv h^H \mv Q \mv h$ as the WD's energy harvesting constraint  for the orthogonal case.
  Moreover, since the WET does not generate interference  to the IR  for the orthogonal case,  the signal-to-noise-ratio at the IR is $\left|\mv g^H \mv w\right|^2 / \sigma^2$, and thus the WD's achievable data rate  is   $R_o=(1-\tau)\log_2\left(1+ \left|\mv g^H \mv w\right|^2 / \sigma^2\right)$.
We thereby formulate the data rate maximization problem in this scheme as \cite{Liu.TWC.14}
\begin{align}
{\rm (P5)}:~\max_{\tau\in(0,1),\mv Q \succeq \mv 0,\mv w} ~&(1-\tau)\log_2\left(1+ \left|\mv g^H \mv w\right|^2 / \sigma^2\right)  
 \nonumber \\ 
 \mathrm{s.t.}~&\textrm{(1)~and}~\!\left\|\mv w\right\|^2 \!\!\le\! \eta \frac{\tau}{1\!-\!\tau} \mv h^H \mv Q \mv h. \nonumber
\end{align}

The following proposition gives the optimal   $\mv Q$,  $\mv w$, and $\tau$ to problem (P5), which are denoted by $\mv Q_o^{*}$,  $\mv w_o^{*}$, and $\tau^*$, respectively.
\begin{proposition} \label{proposition: P2}
 The optimal solutions to  problem (P5) are given by $\mv Q_o^{*}\!=\!\frac{P \mv h \mv h^H}{ \|\mv h\|^2}$,   $\mv w_o^{*}\!= \! \sqrt{\eta \frac{\tau^{*}}{1-\tau^{*}} \mv h^H \mv Q_o^* \mv h} \frac{\mv g}{\|\mv g\|} $, and  $\tau^{*}$ being the unique solution to $\frac{df(\tau)}{d\tau}=0$, where
$
  f(\tau) \triangleq (1-\tau) \log_2(1+\kappa \tau/(1-\tau))
$
 with  $\kappa=\frac{\|\mv g\|^2 \eta}{\sigma^2} \mv h^H \mv Q_o^* \mv h$.
\end{proposition}

Proposition \ref{proposition: P2} can be easily verified by first deriving  the optimal $\mv Q$  and $\mv w$ under any given $\tau$ based on a similar method used to solve (P1) in Section III, and then finding the optimal $\tau$ by solving a simple single-variable convex optimization problem. The proof is thus omitted here for brevity.

Note that among the three reference schemes, the two co-channel schemes can be implemented at the ET and the WD in a distributed manner, while the time-division based orthogonal scheme requires the ET and the WD to coordinate in deciding the optimal time division ratio $\tau^*$.

\section{Numerical Results}
This section provides numerical results  to validate our studies.
We use a similar channel model as  in \cite{Sun.letter.15} and \cite{Xu.letter.15}, by considering Rican fading   for the WET link, and Rayleigh fading  for the WIT and interference links.
We     set   $N_E\!=\!N_W\!=\!2$, $\eta\!=\!0.4$, and   $\sigma^2\!=\!-90$ dBm.  As in \cite{Zeng.letter.15}, we model the loop channel as $\mv \varphi\!=\!\sqrt{\beta}[1~1]^{T}$ with $\beta\!=\!-15$  dB.

 \begin{figure}[t]
\centering
\DeclareGraphicsExtensions{.eps,.mps,.pdf,.jpg,.png}
\DeclareGraphicsRule{*}{eps}{*}{}
\includegraphics[angle=0,width=0.7 \textwidth]{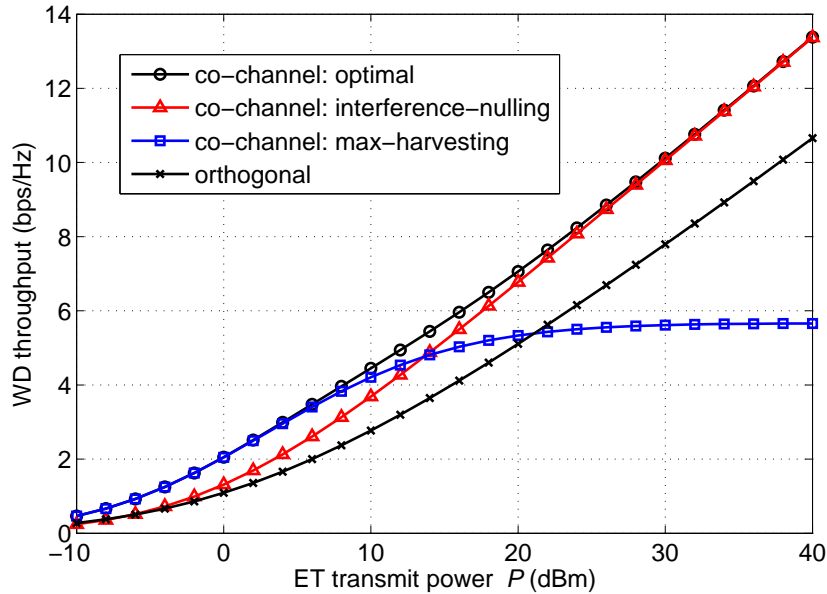}
\caption{WD's throughput versus ET's transmit power.}
\label{fig: benchmark}
\end{figure}

Fig.~\ref{fig: benchmark} compares  the WD's throughput achieved by different transmission schemes versus the transmit power $P$ at the ET.    It is observed  that the throughput achieved by the optimal co-channel scheme is always higher than that achieved by the two reference co-channel schemes.
In particular, when $P$ is sufficient large (or small), the throughput under the reference co-channel scheme with interference nulling (or harvested energy maximization) is observed to be identical to the optimal co-channel scheme.   This is consistent with Remark 4.1 and Remark 4.2.
Moreover, it is observed that as compared to the orthogonal scheme,  a   large throughput gain is  achieved by our proposed co-channel design,  since both energy and information transfer are efficiently operated across the entire time block with carefully controlled co-channel interference to achieve higher spectral efficiency.

\section{Conclusion}
This letter proposed a new  co-channel energy and information transfer scheme for the WPC system with separated ET and IR. By considering multiple antennas at the ET and the WD, we  jointly designed the
ET's  energy beamforming and WD's information beamforming, so as  to maximize the WD's throughput subject to the individual transmit power constraints at the ET and the WD. It was  shown that the optimal transmitted energy signal can properly balance ET's  energy transfer efficiency to the  WD  and its resultant interference to the IR. It was  also revealed that the optimal joint design can be implemented at the ET and WD in a distributed manner. It is our hope that this work can provide insights on the joint WET and WIT design in WPC systems with co-channel energy and information transfer, and it will be an interesting future direction to extend this work to multi-user WPC networks with co-channel energy and information transfer.

\appendices
\section{Proof to Proposition~\ref{proposition: optimal_Q}}
As the  throughput $R$ monotonically increases with the  SINR $\gamma$, by omitting the constant $\eta\|\tilde{\mv g}\|^2$ in $\gamma$, optimizing $\mv Q$ in problem (P3) is equivalent to solving the following problem.
\begin{align}
{\textrm{(P3-1)}}:~\max_{\mv Q\succeq \mv 0} ~ &\frac{\mv h^H \mv Q \mv h}{\mv f^H\mv Q \mv f + \sigma^2}, \nonumber \\
~ \mathrm{s.t.}~& \textrm{(1)}.
\end{align}
It is easy to verify that the optimal solution to problem (P3-1) is achieved with the constraint (\ref{eq: ET_constraint}) being met with equality.
Accordingly,  by substituting $\mathrm{tr}(\mv Q)/ P=1$ into (P3-1) and omitting the constraint in (P3-1),  we have
\begin{align}
\!\!{\textrm{(P3-2)}}\!:\max_{\mv Q\succeq \mv 0}  \hat{\gamma}\!\triangleq\! \frac{\mv h^H \mv Q \mv h}{\mv f^H\!\mv Q \mv f \!\!+ \!\frac{\sigma^2\mathrm{tr}(\mv Q)}{P}}\!=\!\frac{\mv h^H \mv Q \mv h}{\mathrm{tr}\!\left(\!\left(\mv f^H\! \!\mv f \!\!+\!\frac{\sigma^2}{P}\!\mv I\!\right)\! \mv Q \!\right)}. \label{gamma}
\end{align}
Since any feasible solution to (P3-1) is also feasible for (P3-2), optimal value achieved by  (P3-2) is indeed an upper bound of that by (P3-1). Thus, if  an optimal solution to   (P3-2) is feasible to (P3-1),  it is also the optimal solution to (P3-1). Based on such an observation, we can solve (P3-1) by finding an optimal solution to (P3-2) that is feasible to (P3-1).

To solve (P3-2), let  $\hat{\mv Q}\! \triangleq \! \left( \mv f \mv f^H\!\!+\!\frac{\sigma^2}{P}\mv I\right)^{\!\!1/2\!}\!\! \mv Q\! \left( \mv f \mv f^H\!\!+\!\frac{\sigma^2}{P}\mv I\right)^{\!\!1/2}$, and thus  $\mv Q \!= \!\left( \mv f \mv f^H+\frac{\sigma^2}{P}\mv I\right)^{\!\!-1/2}\!\!\hat{\mv Q}\! \left( \mv f \mv f^H+\frac{\sigma^2}{P}\mv I\right)^{\!\!-1/2}$ and $\mathrm{tr}\left( \left( \mv f \mv f^H\!\!+\!\frac{\sigma^2}{P}\mv I\right)\! \mv Q\right)\!=\!\mathrm{tr}\left(\hat{\mv Q} \right)$. By replacing $\mv Q$ with  $\hat{\mv Q}$, we   rewrite  $\hat{\gamma}$ in (\ref{gamma}) as
\begin{align}
\hat{\gamma}
=\frac{\mv h ^H \!\!\left( \mv f \mv f^H\!\!+\!\frac{\sigma^2}{P}\mv I\right)^{\!\!\!-1/2\!\!}\hat{\mv Q}\! \left( \mv f \mv f^H\!\!+\!\frac{\sigma^2}{P}\mv I\right)^{\!\!\!-1/2\!\!}\mv h}{\mathrm{tr}\left(\hat{\mv Q} \right)}.\label{eqn:x}
\end{align}
It is noted that the term in (\ref{eqn:x}) is a Rayleigh quotient. Denote $\hat{\mv Q}^* \!= \!\left( \mv f \mv f^H+\frac{\sigma^2}{P}\mv I\right)^{\!\!-1/2}\!\mv h \mv h^H \!\left( \mv f \mv f^H+\frac{\sigma^2}{P}\mv I\right)^{\!\!-1/2}$. Thus, any positively scaled matrix of $\hat{\mv Q}^*$ is optimal to maximize $\hat{\gamma}$ in (\ref{eqn:x}).
Accordingly, any positively scaled matrix of  $\mv Q^{\star} =  \left( \mv f \mv f^H+\frac{\sigma^2}{P}\mv I\right)^{-1}\mv h \mv h^H \left( \mv f \mv f^H+\frac{\sigma^2}{P}\mv I\right)^{-1} $ is optimal to (P3-2).
By using this result together with $P\big/\mathrm{tr}(\mv Q) = 1$ for (P3-1),  it is then easy to obtain
$\mv Q^{*}\!=\!P\mv Q^{\star}\big/ \mathrm{tr}\left(\mv Q^{\star}\right)\!=\!P \mv v \mv v^H \big/ \|\mv v \|^2.$ is optimal for (P3-2) and is feasible for (P3-1), and thus is optimal for (P3-1).
Proposition \ref{proposition: optimal_Q} thus follows.

\end{document}